\documentclass[aps,groupedaddress,showpacs,showkeys]{revtex4-1}

\usepackage{bm}
\usepackage{graphicx}
\usepackage{gensymb}
 \usepackage{epsfig}
\usepackage{epstopdf}
\usepackage{amsmath}
\usepackage{subfigure}
\usepackage{lipsum}
\RequirePackage[colorlinks,citecolor=blue,urlcolor=blue,linkcolor=blue]{hyperref}
\usepackage{dcolumn}
\begin{document} 


\title{CrCTe$_3$: Computational design of a robust two-dimensional anti-ferromagnetic semiconductor}
\author{Satyananda Chabungbam}
\author{Prasenjit Sen}
\affiliation{Harish-Chandra Research Institute, HBNI, Chhatnag Road, Jhunsi, Allahabad-211 019, India.}






\begin{abstract}
Using density functional theory calculations we establish the hitherto unknown compound CrCTe$_3$ 
to be a stable anti-ferromagnetic 
semiconductor in the R$\bar{3}$ crystal structure with an indirect fundamental gap . Successive layers 
in the bulk compound are weakly bound by
van der Waals forces so that individual layers can be easily exfoliated. A monolayer of CrCTe$_3$ is also an anti-ferromagnetic 
semiconductor. The monolayer is structurally stable over a large range of compressive and tensile strains, and the 
anti-ferromagnetic state is robust over this strain range. Band gap of the monolayer can be tuned by as much as 50\% by
applying strain in this range.

\end{abstract}


\maketitle

\section{Introduction}

Since the exfoliation of graphene sheets from graphite~\cite{Novoselov-2004} little more than a decade ago, two dimensional (2D) materials have
attracted a lot of research attention. A major motivation for this is use of these materials in electronic applications which may open
up new avenues not available to 3D materials. Many interesting 2D materials have been proposed in the past few years which
include compounds such as hexagonal boron nitride (h-BN), transition metal dichalcogenides, and elemental 
2D materials such as phosphorene, germanene, silicene, stanene, arsenene, borophene and
aluminene~\cite{Giem-2007, Hwang-2007, Yafei-2008, Mak-2010, Andrea-2010, Han-2014, Likai-2014, Davila-2014, 
Caha-2009, Mannix-2015, Zhu-2015, Kamal-2015, Kamal-2015b}. Many of these have been realized in experiments
in 2D single layer form (transition metal dichalcogenides, for example), for some, multilayers have been created (arsenene)~\cite{Tsai-2016},
while some others remain as theoretical proposals as yet (aluminine)~\cite{Kamal-2015b}. 
Some of these are promising for applications such as field effect transistors (FET). MoS$_2$ and phosphorene are
good examples of this. Even though there are reports on intrinsic magnetism in MoS$_2$ and WS$_2$ 
arising out of zig-zag edges in the grain boundaries and defects~\cite{Tongay-2012,Mao-2013}, most of these 2D 
materials are non-magnetic, and hence are not suitable for spintronics or storage applications.
There can be two ways of making 2D materials magnetic: functionalizing them by incorporating 
magnetic elements (transition metal (TM) atoms, for example), or identifying novel 2D materials that are inherently magnetic.

Search for magnetic 2D materials has generated new interest in the class of TM tri-chalcogenides with the general formula MAX$_3$ 
(M= $3d$ TM  atom, A=group 14 or 15 element, and X=chalcogen atom). Bulk compounds in this family have been known for
quite some time~\cite{Kurita-1988}. But it is only recently that mono- and few layers of these compounds have been successfully exfoliated~\cite{Du-2016}.
One major advantage of the MAX$_3$ family is that it has a very wide range of chemical diversity as well as structural 
complexity. Compounds in the MAX$_3$ family display interesting properties such as half-metallicity, or semiconducting coupled 
with ferromagnetic (FM) and anti-ferromagnetic (AFM) order. Though MAX$_3$ compounds have been studied extensively 
in the past~\cite{Kurita-1988, Siber-1996, Carteaux-1995-2}, recently the Si and Ge compounds 
CrSiTe$_3$ and CrGeTe$_3$ have attracted a lot of renewed attention~\cite{Chen-2015,Williams-2015,Sivadas-2015,Lin-2016,Gong-2017}.

FM materials  are widely used in spintronics and their behavior is also well understood.  
AFM materials, on the other hand,  are less explored due to their complex magnetic structures. 
In spite of their complexity,  AFM materials have generated a lot of interest 
recently~\cite{Hals-2011, Tveten-2013, Gomonay-2013, Shick-2010, Loth-2012, Manfred-2008}. 
AFM-ordered nano-structures consisting 
of small amounts of Fe-atoms have been demonstrated to encode and store information using a spin-polarized tunneling current~\cite{Loth-2012}. 
A novel phenomenon related to AFM materials is the exchange bias effect~\cite{Meikle-1957}, which is being used for development of 
magnetic read heads in hard disks and magnetic memory devices. FMs can create parasitic magnetic fields that 
can interfere with each other. AFMs do not create such redundant magnetic fields and are largely insensitive to the effect of 
external magnetic field due to their zero (or nearly zero) total magnetization. Another advantage of AFMs over FMs is that switching 
between different states of an AFM is much faster than that in FM~\cite{Manfred-2008}. So, AFM materials are potential candidates for 
developing ultrafast and ultrahigh density spintronic and storage devices.
Recent experiments show that bimetallic 
antiferromagnets like Mn$_2$Au~\cite{Barthem-2013} and MnIr~\cite{Selte-1968} are suitable for application in tunnel anisotropic 
magnetoresistance (TAMR) devices due to their high magnetic ordering temperatures and strong magnetic anisotropy. 
So, it is interesting and necessary to explore new potential AFM materials that can be used in nano-scale devices.

Both CrSiTe$_3$ and CrGeTe$_3$ are FM semiconductors.
FM ordering in these compounds arises due to Cr-Te-Cr superexchange~\cite{casto-2015}, and has been rationalized
within the Goodenough-Kanamori-Anderson (GKA) rules~\cite{gka}. The Cr atom in these compounds are at the
center of a Te$_6$ trigonal anti-prism with $D_{\rm 3d}$ point group symmetry. The Te$_6$ trigonal 
anti-prism cages are arranged in an edge-sharing manner in each layer of the crystal. 
The Cr-Te-Cr angle is nearly, but not exactly, 90$^0$, and the Cr-Cr distances are relatively large 
($\sim 4$ \AA~ compared to 2.49 \AA~ in bcc Cr), making their direct exchange weak. 
The 90$^0$ cation-anion-cation superexchange
for a $d^3$ configuration of the TM cations in compounds with $O_h$ symmetry is quite complex. 
There are competing FM and AFM couplings.
Generally TM oxides end up having AFM ground states, while the chlorides are FM~\cite{gka}.
The situation is more complicated in the tri-chalcogenides due to two factors: first, the 
crystal-field splitting of the Cr $3d$ levels is slightly different in the $D_{\rm 3d}$ crystal-field 
as compared to the $O_h$ crystal-field assumed in the GKA analyses; second,
the Cr-Te-Cr angle is not exactly 90$^0$, as already stated. Therefore, whether the resultant superexchange is
FM or AFM depends on a fine balance of different factors, and it may be possible to affect the magnetic ground 
state by small structural changes. In fact Casto et al~\cite{casto-2015} have shown that there is a strong spin-lattice
coupling in CrSiTe$_3$.

Compressive in-plane strain 
decreases the distance between the Cr atoms and increases overlap between the occupied $d$ orbitals on neighboring atoms. 
Increase in direct overlap between atomic orbitals may drive the direct exchange
it to be antiferromagnetic beyond a point~\cite{Fazekas}. In bulk bcc Cr, the direct exchange is AFM.
Interestingly, the Si compound is almost at the verge of a FM-AFM transition. 
A mere 0.5\% compressive strain drives 2D layers of CrSiTe$_3$ AFM~\cite{Chen-2015}. 
Ge having a bigger atomic radius, the Cr-Cr distance increases from 3.90~\AA~ in CrSiTe$_3$ to 
3.94~\AA~ in CrGeTe$_3$, and the FM phase is stabilized. The Curie temperature (${\rm T_C}$) increases from 33~K in 
CrSiTe$_3$~\cite{Carteaux-1995,Williams-2015} to 61~K in CrGeTe$_3$~\cite{Carteaux-1995-2}. 
In the theoretically proposed CrSnTe$_3$~\cite{Zhuang-2015}, the Cr-Cr distance increases even further 
(4.05~\AA~ in HSE06)~\cite{Zhuang-2015} and the ${\rm T_C}$ is estimated to be greater than 170~K.

Thus if one could design a similar tri-chalcogenide compound with smaller Cr-Cr distance, 
it is possible that the ground state would be AFM. An obvious option is to replace Si by C. 
Chemically, C would behave the same way as far as crystal binding is concerned. However, being smaller in size, would lead to 
a shorter Cr-Cr distance. The questions are, is the compound CrCTe$_3$ dynamically and mechanically stable in the R$\bar{3}$ crystal
structure? If it is, is the Cr-Cr distance short enough to give rise to an effective AFM superexchange? 
And finally, can 2D layers of this compound be exfoliated? What are the properties of such a novel 2D material?

The rest of the paper is organized as follows. In Section~\ref{comp-details} we describe the theoretical methods used for
our calculations. In Section~\ref{bulk-cct} we discuss stability and electronic properties of bulk CrCTe$_3$ (CCT). In various
subsections of Section~\ref{mcct} we discuss stability and electronic properties of a monolayer CCT. Section~\ref{mcct-strain}
contains our results on strain engineering of a monolayer CCT. Finally, we draw our conclusions in Section~\ref{conclude}.


\section{Computational details}
\label{comp-details}

Spin polarized density functional theory (DFT), as implemented in Vienna Ab-initio Simulation Package 
(VASP)~\cite{Hohenberg-1964, Kohn-1965, Kresse1, Kresse2, Kresse-1999, Kresse3}, was used for all electronic structure calculations. 
Electronic wave functions were expressed in a plane wave basis with an energy cutoff of 500~eV. Electron-ion interactions were treated using the
projector augmented wave (PAW) method. The gradient corrected Perdew-Burke-Ernzerhof (PBE)~\cite{Perdew-1996, Vanderbilt-1990} 
functional was employed for electronic minimization, and structure relaxation. 
It is well known that the local and semi-local exchange-correlation functionals underestimate band gaps. Therefore,
for a more accurate estimation of the band gaps in monolayer CCT, 
the hybrid functional proposed by Heyd, Scuseria and Ernzerhof (HSE06)~\cite{Heyd-2003,Heyd-2006} was used.
Because HSE06 calculations are considerably more expensive than those with local/semi-local functionals, we did not
employ it in any other case.
To account for dispersion interactions, a non-local correlation functional~\cite{Kyuho-2010,Dion-2004,Roman-2009} 
(vdW-DF2) was used. 
Earlier works on graphite have shown that the best description of dispersion interactions can be obtained by the vdW-DF2
functional~\cite{Mapasha-2012,Singh-2014}.
The convergence criteria for electronic minimization and structure optimization were set to 10$^{-6}$~eV and 0.001~eV/\AA~ respectively.
The Brillouin zone (BZ) integration
was performed within the Monkhorst-Pack scheme using $9\times9\times3$ and $11\times11\times1$ meshes for bulk 
and monolayer respectively. A unit cell was used for bulk electronic structure calculations. Bigger supercells were
used for calculating phonon band structure and cleavage energy.
For the monolayer, the usual repeated slab geometry was used with a vacuum space of 20~\AA.  
Phonon band structures were calculated  using the inter-atomic force constants obtained from
VASP within the density functional perturbation theory (DFPT) framework, with the PHONOPY code~\cite{phonopy}
as the post-processor.
Energy and force convergence criteria were set to 10$^{-8}$~eV and 0.00001~eV/\AA~ for these calculations.
A ($3\times3\times2$) supercell was used for bulk phonon calculations. For phonon calculations of monolayers, 
a ($3\times3 $) planar supercell was used.
Visualization packages VESTA~\cite{VESTA} and VSim~\cite{V_SIM} 
have been used to reproduce the crystal structures and vibration modes shown in this work. 

\begin{figure}[ht]
 \includegraphics[scale=0.4]{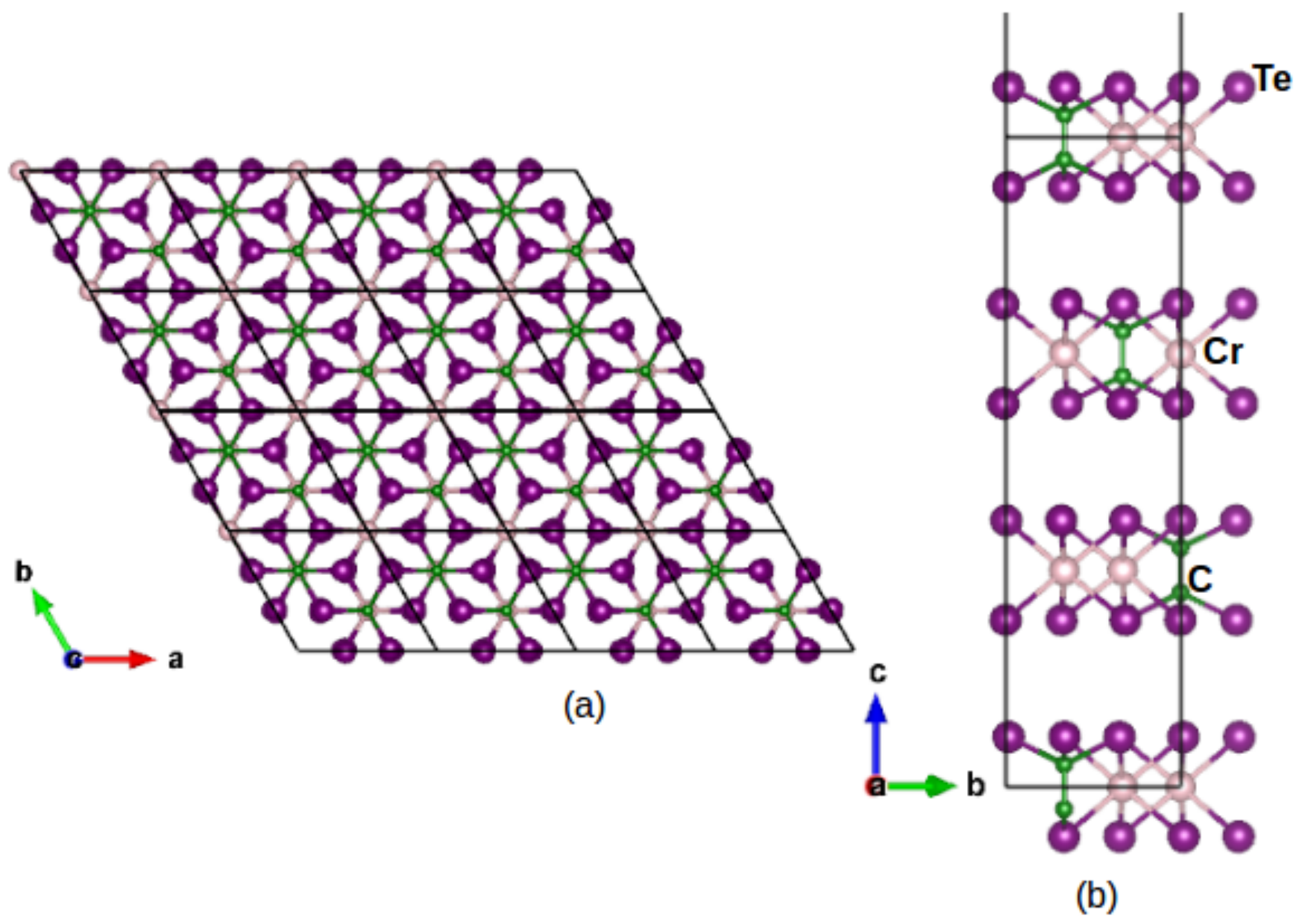}
 \caption{Optimized lattice structure of CrCTe$_3$ (a) top view (b) side view.}  
 \label{fig:bulk-strct}
\end{figure}

\begin{figure}
\begin{center} 
\includegraphics[scale=0.2]{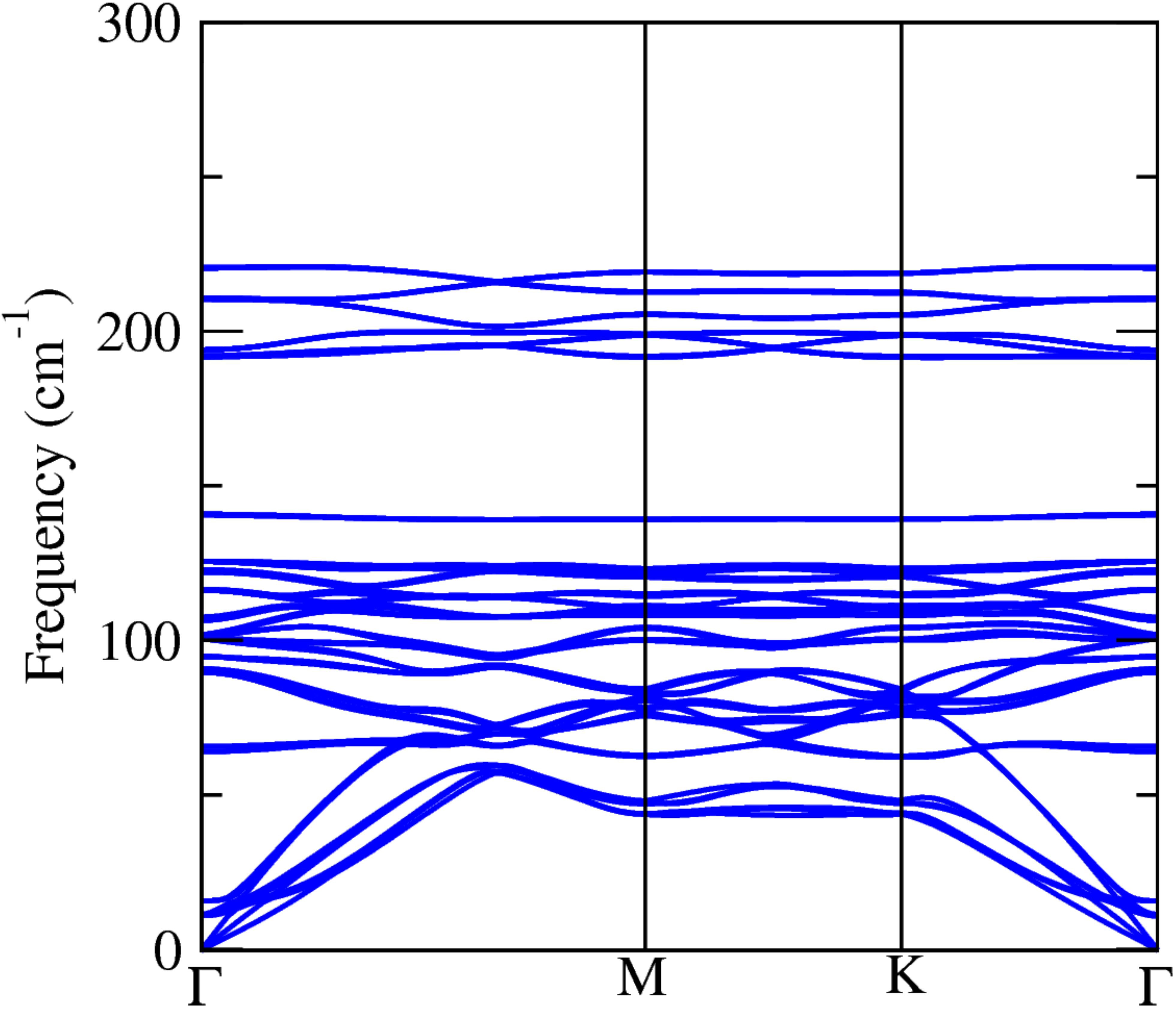}
\end{center}
\caption{Phonon band structure of bulk CrCTe$_3$}
\label{fig:bulk-phon}
\end{figure}

\begin{figure}
\begin{center} 
\includegraphics[scale=0.48]{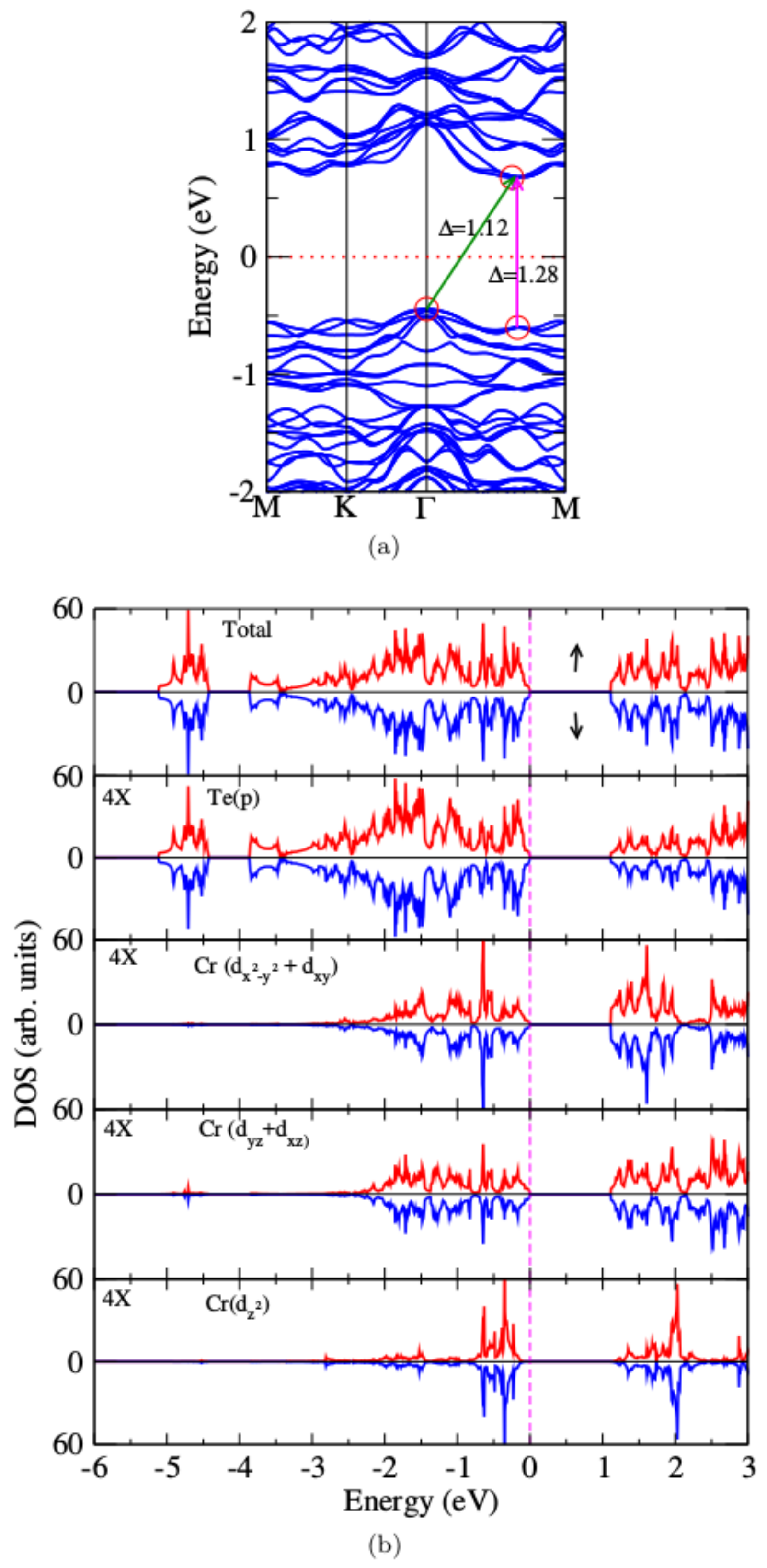}
 \end{center} 
\caption{ (a) Band structure of bulk CrCTe$_3$ (b) Total and orbital resolved DOS of bulk CrCTe$_3$.}  

\label{fig:bands-dos}
\end{figure}


\section{Bulk CCT}
\label{bulk-cct}
\subsection{Structure and magnetic ordering}

Since CrSiTe$_3$, CrGeTe$_3$~\cite{Williams-2015, Carteaux-1995-2}, and the theoretically 
proposed compound CrSnTe$_3$~\cite{Zhuang-2015} all have stable R$\bar{3}$ structure, . 
we assume that CCT also has the same structure. Fig.~\ref{fig:bulk-strct} shows the structure of CCT. 
Each Cr atom is six-fold coordinated to Te-atoms arranged in a trigonal anti-prism structure with $D_{3d}$ point group
symmetry. Each C atom is three-fold coordinated to Te-atoms.
The formal valence of the atoms are Cr$^{+3}$, C$^{3+}$ and Te$^{-2}$, which gives a $3d^3$ electronic configuration
on Cr. Hund's rule would then give spin $S=3/2$ on each Cr atom. We considered 
FM and four different AFM order of the Cr spins. These AFM spin arrangements are shown in Fig S1(a) in
Supplementary Information (SI)~\cite{si}, and their energies are given in Table S1. 
The AFM state in which nearest neighbor Cr spins are oriented in opposite
directions within each CrCTe$_3$ layer and between layers (3D N\'eel order, called AFM1 here) turns out to have the
lowest energy.  The energy difference $\Delta E = E_{\rm FM}-E_{\rm AFM1} = 13.6$~meV/atom.
The magnetic moment comes mostly from the Cr atoms. In the FM state, each Cr atom contributes 3.03 $\mu_B$,
while in the AFM1 state the two Cr atoms in the unit cell contribute $\pm 2.85$ $\mu_B$. Further details about
magnetic moment can be found in Table S3 in SI~\cite{si}. 
This confirms that bulk CCT has an AFM order in its ground state. 
The PBE-vdW-DF2 optimized lattice parameters for 
antiferromagnetic bulk CCT are a = b = 6.64~\AA~ and c = 21.41~\AA~ with an interlayer separation of 3.85~\AA.  
The Cr-Te bond length is 2.80~\AA, and Cr-Cr bond length is 3.83~\AA~, with a Cr-Te-Cr bond angle of 86.24$^{\degree}$.

\subsection{Stability and electronic structure}

While studying a new material theoretically, it is essential to check whether it is dynamically and mechanically stable. In order to
ensure that the optimized R$\bar{3}$ lattice structure for CCT is dynamically stable, we calculated its phonon band 
structure. Any structural instability would show up as soft phonon modes with imaginary frequencies. 
As can be seen in Fig.~\ref{fig:bulk-phon}, all phonon branches have real frequencies signifying dynamical 
stability of the R$\bar{3}$ crystal structure. 
We also calculated the elastic constants $C_{11}$ and $C_{12}$ to test for mechanical stability of CCT. These
elastic constants are defined as, 
\begin{equation}
C_{11} = \frac{1}{V_0}.\frac{\partial^2 E}{\partial \epsilon_{11}^2} \;\; {\rm and} \;\; 
C_{12} = \frac{1}{V_0}.\frac{\partial^2 E}{\partial \epsilon_{11} \partial \epsilon_{12}}.
\end{equation}

\noindent Here $E$ is the total energy of CCT, $V_0$ is its equilibrium volume, and $\epsilon$ are the
components of strain. $C_{11}$ and $C_{12}$ were found to be
112~GPa and 23.50~GPa respectively. The elastic constants satisfy the Born stability criterion 
$C_{11} - C_{12} > 0$ indicating mechanical stability of bulk CCT.

After having established that bulk CCT is structurally stable, we now calculate its electronic structure.
Electronic band structure calculated using PBE-vdW-DF2 is shown in Fig.~\ref{fig:bands-dos}(a).
The valence band maximum (VBM) occurs at the $\Gamma$-point while the conduction band minimum (CBM) occurs
between $\Gamma$ and M-points ($\Delta$-point henceforth). The smallest direct gap appears at the 
$\Delta$-point and is 1.28~eV in PBE. But the fundamental
band gap is an indirect one from $\Gamma$ to $\Delta$, and is 1.12~eV. 
Fig.~\ref{fig:bands-dos}(b) shows the total and orbital resolved density of states (DOS) of bulk CCT. 
C states have very little contribution near the band edges, so contributions of only the Cr and Te atoms 
have been shown.  In the $D_{3d}$ crystal field of the Te anions, the Cr $3d$ states are split into three 
levels: E$_{\rm g}$($d_{x^2-y^2}$, $d_{xy}$); E$_{\rm g}$($d_{yz}$, $d_{zx}$); and A$_{\rm 1g}$($d_{z^2}$) 
as seen in the bottom three panels of  Fig.~\ref{fig:bands-dos}(b). The states near the 
VBM region have contributions from both Te-$p$ and Cr-$d$ orbitals while the states near the CBM have major 
contributions only from the Cr-$d$ states. 

To conclude this section we note that our guess that a shorter Cr-Cr distance in CrCTe$_3$ 
may lead to an AFM ground state for the material is indeed borne out by DFT calculations.
The material is found to be dynamically and mechanically stable, so it should be possible to
synthesize it in the laboratory.

 \section{Monolayer CCT}
 \label{mcct}

  \begin{figure}
   
\includegraphics[scale=0.3]{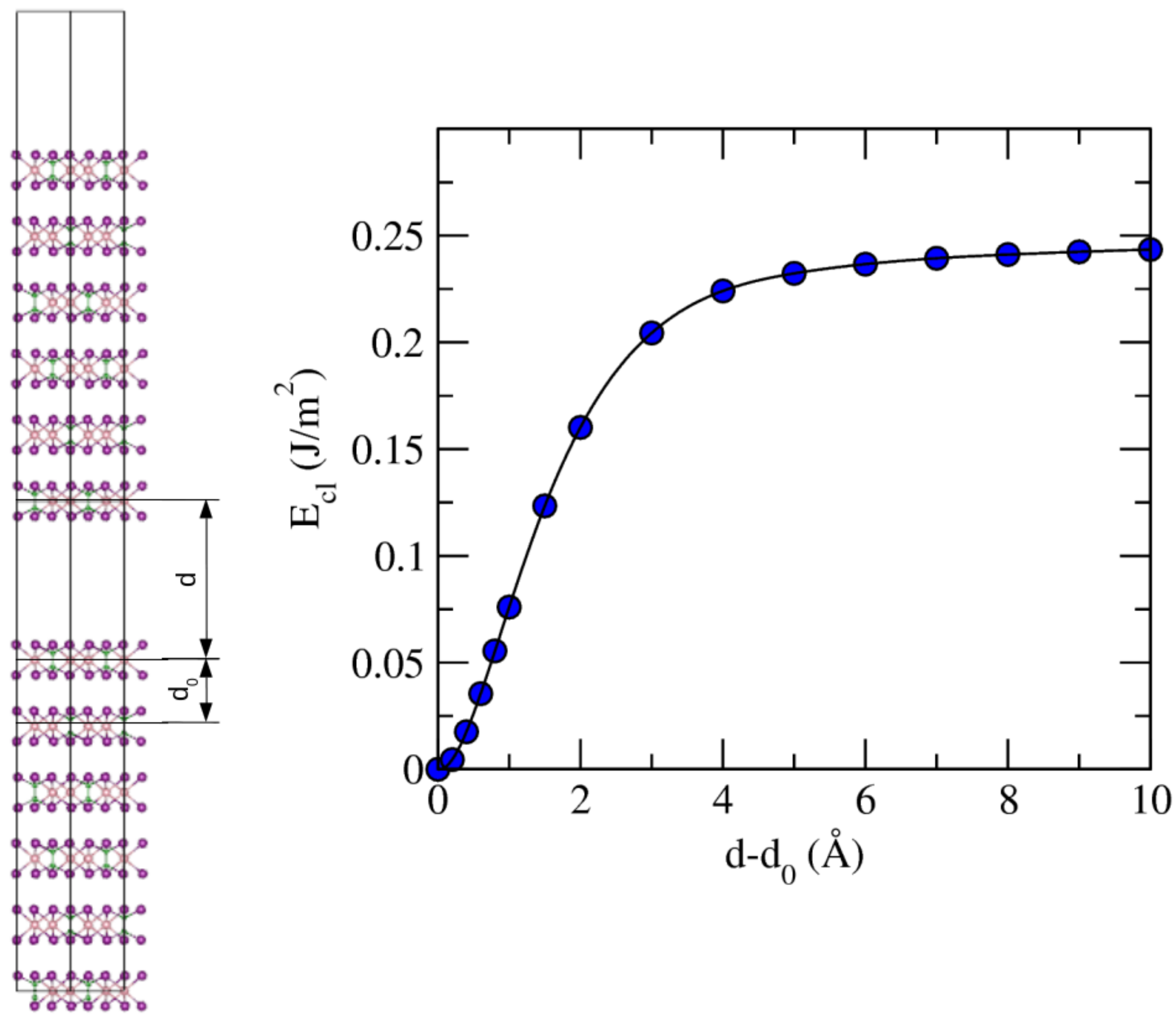}
\caption{Increase in energy as function of distance (relative to the equilibrium separation) 
between two half crystals of CrCTe$_3$.}
  \label{fig:cleavage}
\end{figure}    

We now study properties of 2D monolayers of CCT (MCCT).
Before studying properties MCCT, it is important to find out how easy or difficult it is to exfoliate
such layers from the bulk. 
We estimate the ease of exfoliation by the so-called cleavage energy. Cleavage energy is defined as the
energy required to separate the crystal into two halves along the gap between two successive CCT layers, 
and is calculated as follows. The increase
in energy is calculated as the distance between two halves of bulk CCT 
is increased compared to the equilibrium separation. We have calculated this quantity with three and six
layers in the simulation cell, and we get the same exfoliation energy in both cases indicating that our
estimation does not suffer from finite size effects.
Energy increase for a six-layer supercell is shown in Fig.~\ref{fig:cleavage}. The
energy increases sharply at first, but saturates at larger separations. The saturation value of the energy, 
0.24~J/m$^2$, is the cleavage energy. Interestingly, cleavage energy of CCT is smaller than that of graphite, 
0.37~J/m$^2$~\cite{Wang-2015}. It may be noted that the cleavage energies of other members of the 
family, CrSiTe$_3$, CrGeTe$_3$, are also higher, 0.35~J/m$^2$ and 
0.38~J/m$^2$~\cite{Xingxing-2014} respectively. Since graphene, and CrSiTe$_3$ and CrGeTe$_3$ layers can be easily 
exfoliated using simple mechanical means, one expects that the same procedure would work for CCT. 

\begin{figure}
\includegraphics[scale=0.5]{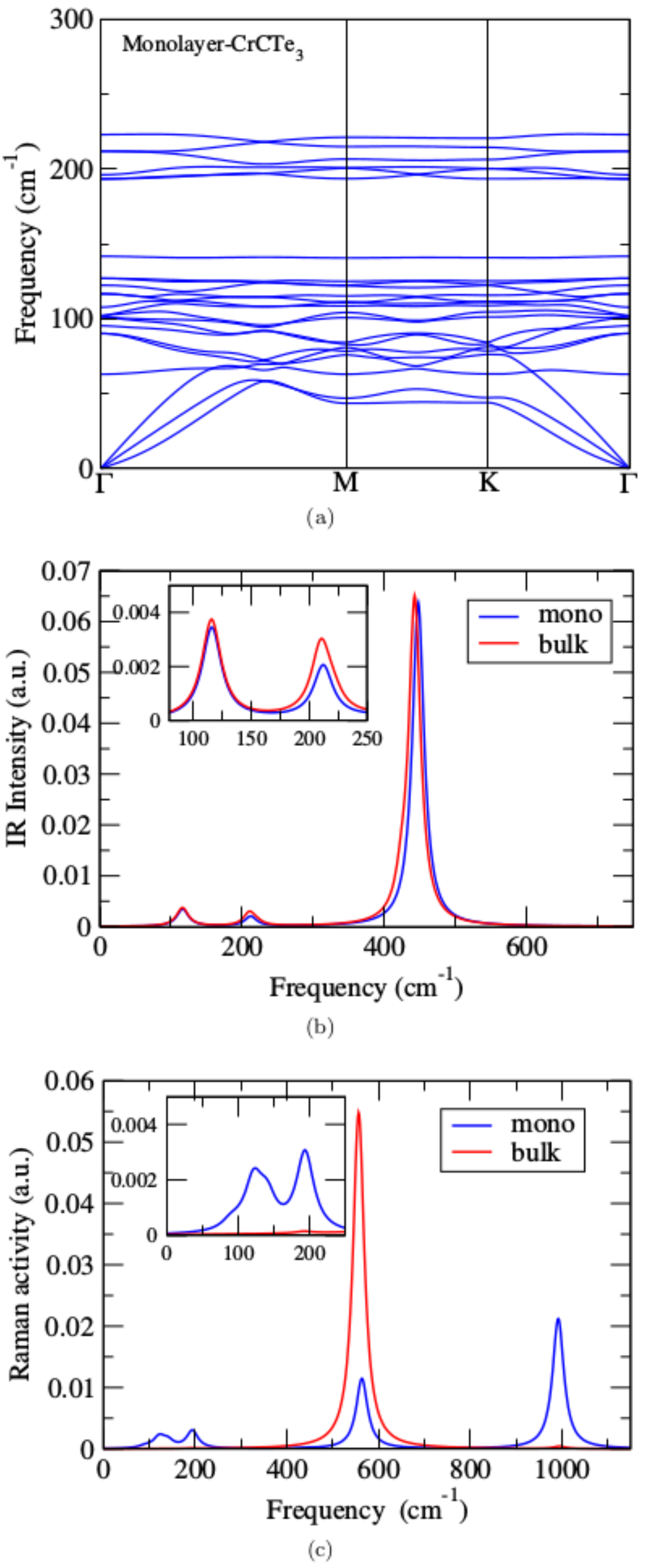}

\caption{(a) Phonon band structure of monolayer CrCTe$_3$; calculated (b) infrared and (c) Raman spectra of monolayer and bulk CCT.
Insets in (b) and (c) are enlarged views of the low-frequency region.}  

\label{fig:monolayer-phonon-raman}
\end{figure}

\subsection{Stability}

Before calculating its electronic properties, we re-optimized the in-plane lattice parameters for MCCT should 
they change compared to the bulk values.
The PBE-vdW-DF2 optimized lattice parameters for monolayer CCT turn out to be $a = b = 6.63$~\AA, very close to the
bulk values. Such an insignificant change in the lattice constants is perhaps because of a relatively weak inter-layer binding.
In monolayer CCT we calculated energies of FM and three different AFM ordering of the Cr spins:
the 2D N\'eel, zigzag and stripe phases (Fig.\ S1(b) in the SI~\cite{si}).
Similar to its bulk form, the 2D N\'eel state (AFM1) turns out to have the lowest energy with $\Delta E = $14.6~meV/atom.
Energies of all the magnetic states are given in Table S2 in the SI~\cite{si}.
The moment on each Cr atom in the FM state in MCCT is practically same as what we found in the bulk, 2.98 $\mu_B$.
The moment on the two Cr atoms in the AFM1 state are $\pm 2.76$ $\mu_B$, slightly lower than the bulk value. 
The energy difference between the FM and AFM1 states in a monolayer is thus very close to that in the bulk,
perhaps because the coupling between successive layers is rather weak.

From our calculations of the FM and AFM1 states, we estimated the N\'eel temperature ($T_{\rm N}$) of the monolayer by treating
the Cr spins as $S=3/2$ Ising spins. 
The energy difference between the two spin ordered states, gives the nearest neighbor exchange interaction energy
as 10.8 meV. Using the expression for the transition temperature for an Ising model on a honeycomb lattice~\cite{Wannier-1945},
($T^* \sim JS^2/1.3 k_B$) we find $T_{\rm N} \sim 217 K$. 
We have used the fact that the transition temperature in the nearest-neighbor Ising model
is independent of the sign of the exchange interaction.

It is important to check the dynamical and mechanical stability of a monolayer as well. 
As in the case of bulk, phonon band structure is calculated for the 2D layer
and is shown in Fig.~\ref{fig:monolayer-phonon-raman}(a). All phonon branches turn out to have positive frequencies signifying
dynamical stability of the structure. 
Te, being the heaviest of all the constituent atoms, dominates the lower frequency region, which is followed by Cr and C. 
Similar to CrSiTe$_3$~\cite{casto-2015}, flat optical phonon branches is a feature of CrCTe$_3$ also.

To check for mechanical stability of the 2D monolayer, we calculated its elastic constants. 
Elastic constants of a monolayer are defined as
\begin{equation}
C_{11} = \frac{1}{A_0}.\frac{\partial^2 E}{\partial \epsilon_{11}^2} \;\; {\rm and} \;\; 
C_{12} = \frac{1}{A_0}.\frac{\partial^2 E}{\partial \epsilon_{11} \partial \epsilon_{12}}.
\end{equation}

\noindent Here $E$ is the total energy of MCCT, $A_0$ is its equilibrium area, and $\epsilon$'s are the components of strain.
The calculated values of elastic constants for MCCT are: C$_{11}=81.70$ N/m, C$_{12}=17$ N/m which are slightly lower 
than that of bulk CCT. MCCT also satisfies Born's criterion for mechanical stability $C_{11}-C_{12} > 0$. 
To avoid curling during the exfoliation process of a 2D crystal, a high in-plane stiffness is necessary. To estimate the 
in-plane stiffness, we calculate the in-plane Young's modulus of monolayer CCT using the formula: 
$Y_s=(C_{11}^2-C_{12}^2)/C_{11}$. For monolayer CCT, $Y_s$= 78 N/m. This is about 23$\%$ of that of  
graphene (341 N/m)~\cite{Peng-2013}, one of the strongest materials. Thus, it can be assumed
that MCCT can keep its free-standing structure. 
The in-plane stiffness of monolayer CCT is higher than that of monolayer CrSnTe$_3$ with 
$Y_s$= 55 N/m [C$_{11}=60$ N/m, C$_{12}=17$ N/m]~\cite{Zhuang-2015}. 
Therefore we conclude that monolayers of CCT are dynamically and mechanically stable, and
they can be mechanically exfoliated for further studies in their free-standing structures.

2D materials are often characterized by their infra-red (IR) and Raman spectra. Therefore, we have calculated
the IR and Raman spectra for both bulk and a monolayer CCT.
Off-resonant Raman activity of a mode was calculated by computing the derivative of macroscopic dielectric 
tensor with respect to normal mode coordinates~\cite{Porezag-1996}. For this, phonons at 
Gamma, and macroscopic dielectric tensors were calculated using DFPT as implemented in VASP. 
The derivatives were calculated using the script developed by Fonari and Stauffer~\cite{raman}.

For infrared modes, the tensor of the Born effective charges 
(the first derivative of the polarization with respect to the ionic coordinates) was calculated using DFPT.
Within the dipole approximation, the infrared intensity ($I$) of an eigenmode can 
be expressed in terms of the Born effective charges $Z^*_{\alpha \beta}$ and the eigenvectors $e_{\beta}(l)$, 
where $\alpha$ and $\beta$ 
are cartesian polarizations, and $l$ labels the atoms of the system~\cite{Karhanek-2010,Karhanek-script}.
\begin{equation}
I = \sum_{\alpha} \left [ \sum_{l,\beta} Z^*_{\alpha,\beta} e_{\beta}(l) \right ] ^2
\end{equation}

Bulk CCT has six infra-red (IR) active modes. These are at 442.9 cm$^{-1}$, 
210.5 cm$^{-1}$ and 116.3 cm$^{-1}$. All these frequencies are doubly degenerate.
These are shifted to slightly higher energies,
447.8, 212.3 and 116.5 cm$^{-1}$ respectively in MCCT. It may be noted that out of the six IR-active modes, the first peak 
is the most prominent one in both bulk (442.9 cm$^{-1}$) and monolayer CCT (447.8 cm$^{-1}$) 
(Fig.~\ref{fig:monolayer-phonon-raman}(b)). 
Analyzing the eigenvector of these phonon modes at $\Gamma$, it is seen that these 
involve motion of the C atoms.  
The second peak at  210.5 cm$^{-1}$ in bulk (212.3 cm$^{-1}$ in monolayer) 
comes from Te-Cr-Te bond bending. The third one at 116.3 (116.5)~cm$^{-1}$ is from both C-Te 
bond stretching and Te-Cr-Te bond bending. These are very similar to what was found in CrSiTe$_3$~\cite{casto-2015}.

Eight Raman active modes have been found in MCCT. Three of these,
at 992.2, 563.8 and 193.8~cm$^{-1}$, are most intense. These modes are found in bulk CCT too but at 
994.9, 556.7 and 192.3~cm$^{-1}$. (Fig.~\ref{fig:monolayer-phonon-raman}(c)). 
Atomic motions corresponding to the IR and Raman active modes are shown in Figs. S2 and S3 in the SI~\cite{si}. 
Thus the first Raman active mode shifts to a lower energy in the monolayer
while the other two shift to higher energies like the IR-active modes.

 \subsection{Electronic structure} 
\begin{figure}
\includegraphics[scale=0.75]{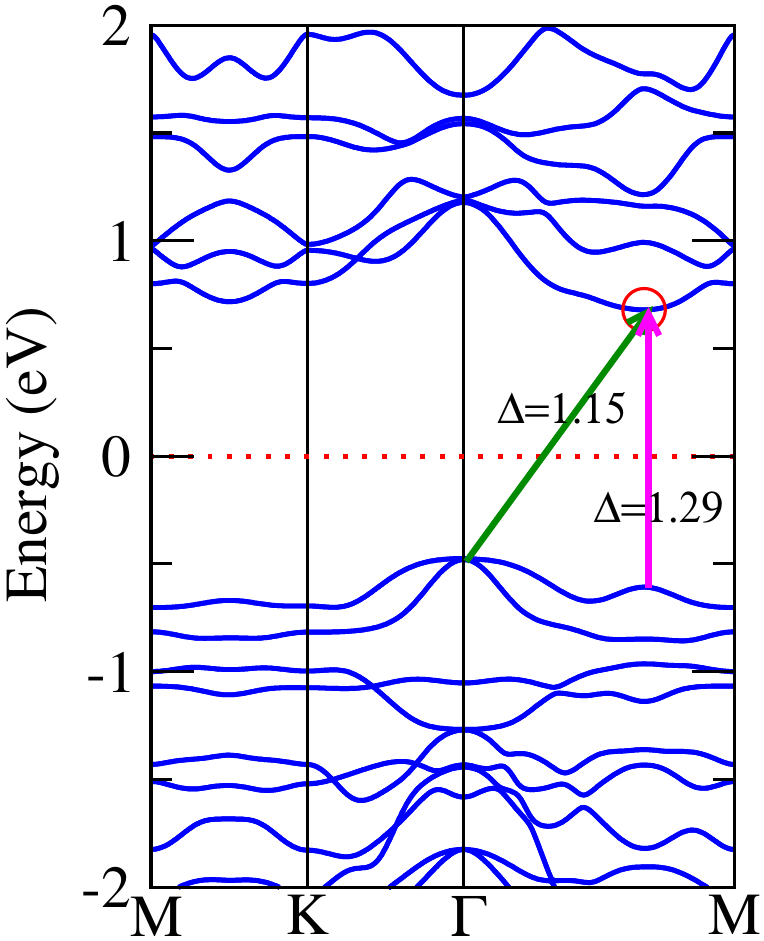}
    \caption{Electronic band structure of monolayer CrCTe$_3$. Direct and indirect band gaps are shown by arrows.}   
\label{fig:monolayer-band}
\end{figure}

\begin{figure}
\includegraphics[scale=0.6]{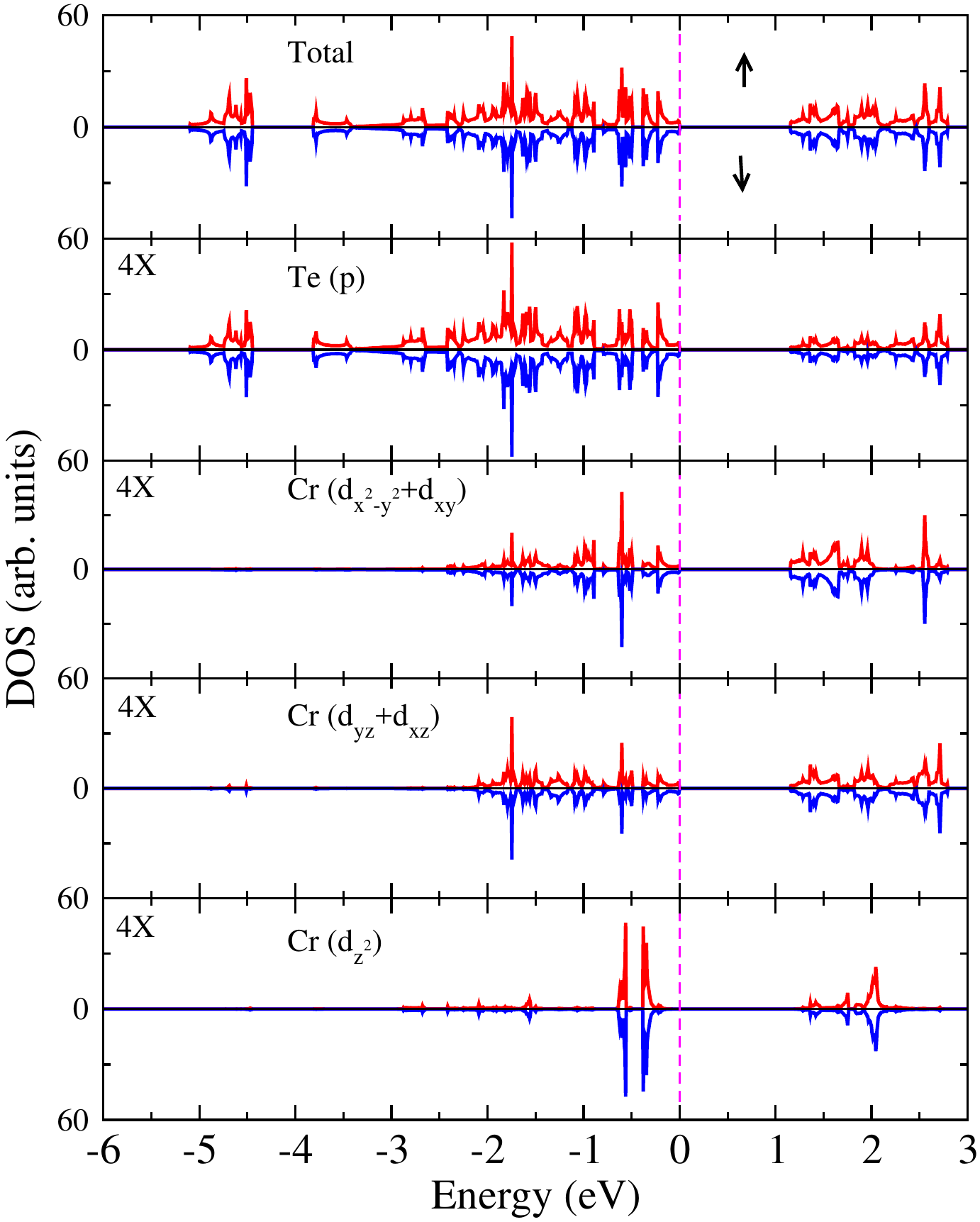}

 \caption{Total and orbital resolved DOS of monolayer CrCTe$_3$. }  
\label{fig:monolayer-dos}
\end{figure}

We now proceed to study the electronic structure of MCCT. 
The electronic band structure of MCCT calculated using PBE-vdW-DF2 is shown in Fig.~\ref{fig:monolayer-band}. 
The band structure near the gap looks very similar to the bulk except that there are fewer bands in the monolayer, 
as there are a smaller number of 
atoms in the unit cell. Again, as the inter-layer coupling is weak, band dispersion arises mainly due to in-plane bonding. Thus there
is little difference between the band structure of the bulk
and the monolayer. As a consequence, the energy gaps in
a monolayer are also nearly the same as those in the bulk.
The smallest direct PBE gap is 1.29 eV at the CBM which is at the $\Delta$-point. The
fundamental band gap is indirect and is equal to 1.15~eV from $\Gamma$ to $\Delta$. 
Band structure of MCCT calculated with the HSE06 functional is given in Fig. S4 of the
SI~\cite{si}. We get qualitatively the same band structure with
larger band gaps. In HSE06, the direct (at $\Delta$) and indirect ($\Gamma$ to $\Delta$)
gaps are nearly the same, and are 1.87~eV. Since the bulk and the monolayer have almost
the same band gap at the PBE-vdW-DF2 level, it is reasonable to believe that the HSE06
functional would produce a similar band gap for the bulk material. Thus a more accurate 
estimate of band gap in the bulk would be around 1.87~eV.
  
To understand contributions of different atomic states to the electronic structure of monolayer CCT, 
we have plotted the total and atom and orbital resolved DOS in Fig.~\ref{fig:monolayer-dos}.
As in the bulk, states near the VBM has contributions from both Te-$p$ and Cr-$d$ (mostly $x^2-y^2$, $xy$,
$yz$ and $zx$) states. CBM has contributions mostly from the Cr-$d$ states.

\begin{figure*}
\includegraphics[scale=0.4]{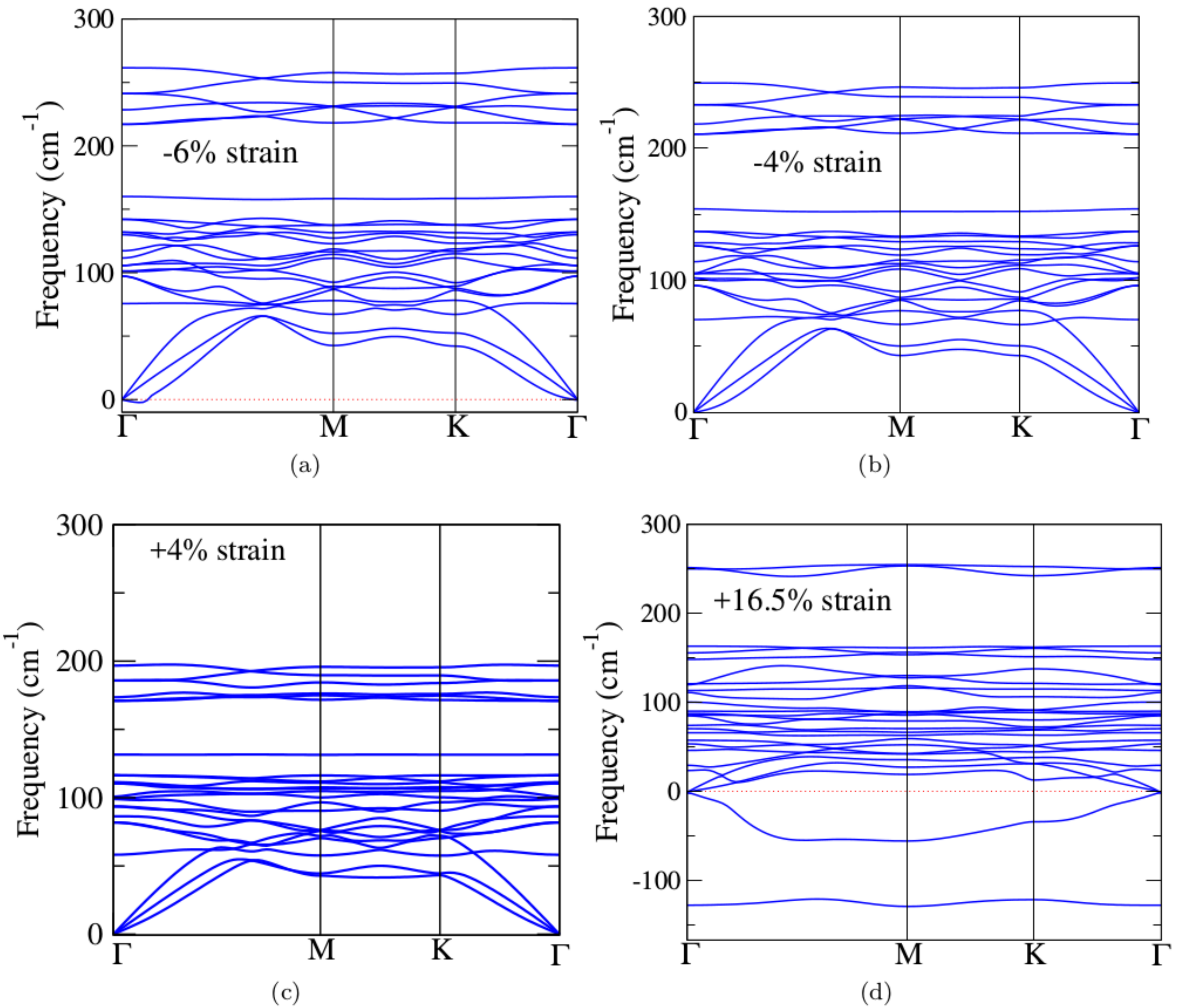}
  \caption{Phonon band structures of monolayer CrCTe$_3$ under different compressive and tensile strain. }    
\label{fig:strain-phonon}
\end{figure*}

\begin{figure*}
\includegraphics[scale=0.38]{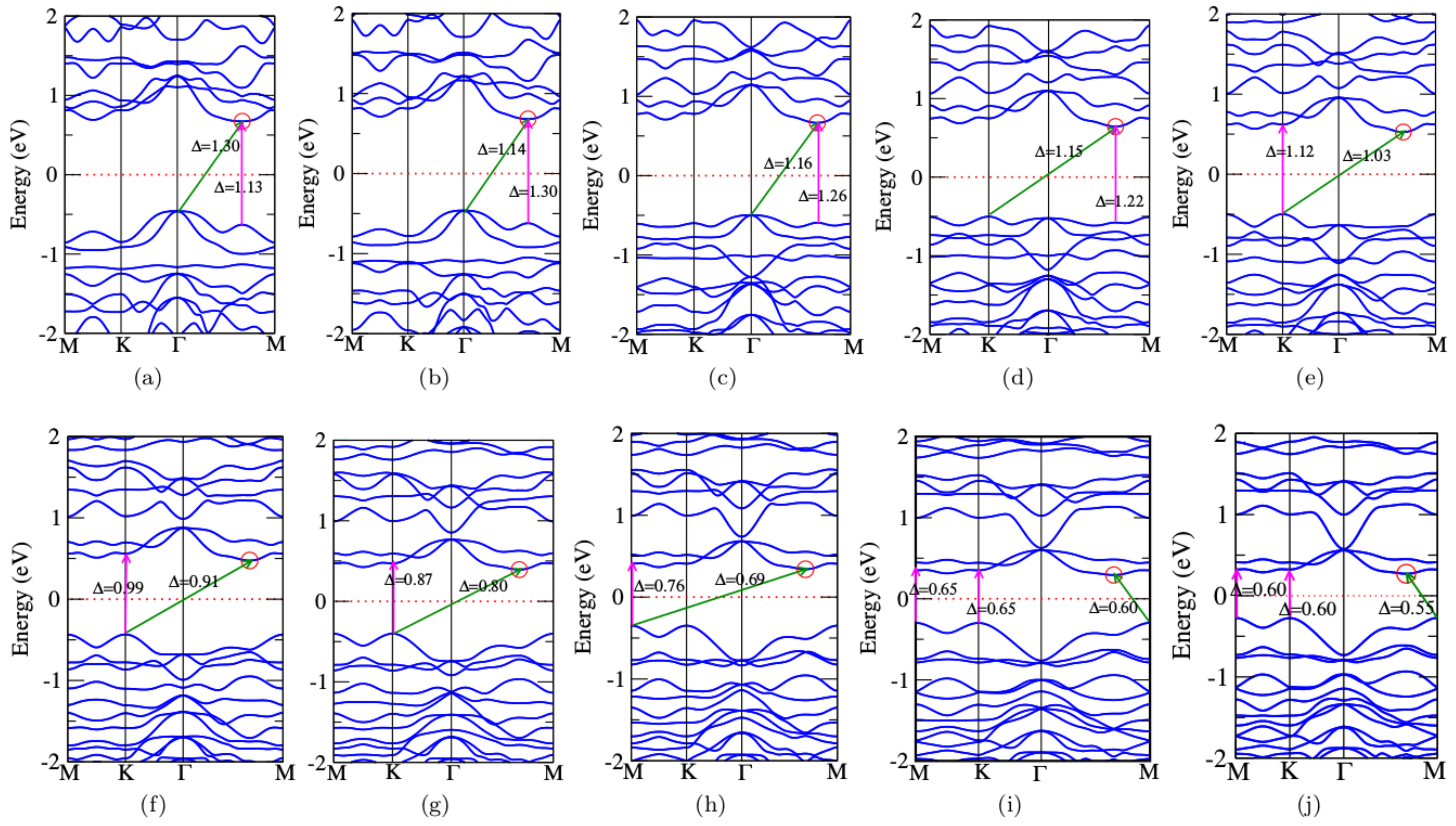}  
         
  \caption{PBE electronic band structure of monolayer CrCTe$_3$ under different strains 
  (a)$-4$\% (b)$-2$\% (c)$+2$\% (d)$+4$\% (e)$+6$\% (f)$+8$\% (g)$+10$\% (h)$+12$\% (i)$+14$\% 
  (j)$+15$\% (direct and indirect gaps are shown by arrows).}    
\label{fig:strain-band}
\end{figure*}

\section{Strain engineering of Monolayer CCT}
\label{mcct-strain}

Next we explore how electronic properties of MCCT can be engineered using in-plane strain. 
In this work we consider application of biaxial strain only. The first point to check is the range of strain over
which the R$\bar{3}$ structure is  dynamically stable. For this, phonon band structure is calculated at
different applied strain. Based on our results, we claim,
with a caveat, that the R$\bar{3}$ structure for MCCT is stable from 
4\% compressive to 16\% tensile strain as all phonon branches have positive frequencies in this range.
Fig.~\ref{fig:strain-phonon} shows the calculated phonon spectra of MCCT 
at different compressive (negative) and tensile (positive) strains. 
Phonons are calculated within a harmonic approximation for the interatomic forces. Whether this 
approximation remains valid up to 16\% strain is a pertinent question. How to apply
such large tensile strain in practice is also not obvious. Therefore, while our results are definitely valid
up to few percent strain, they should be taken as indicative at larger strains. 

A long wavelength acoustic phonon branch becomes unstable at $-5$\% strain. This soft phonon mode
is clearly seen in at Fig.~\ref{fig:strain-phonon}(a) at $-6$\% strain. 
The eigenvector of this mode indicates that this is a flexural mode (Fig.\ S5 in SI)~\cite{si}.
This instability thus indicates a buckling transition in MCCT. It is interesting that the buckling instability sets in
beyond 4\% compressive strain in CCT whereas graphene develops this instability at a much smaller strain of 
0.75\%~\cite{Kumar-2010}. Beyond 16\% tensile strain an acoustic branch becomes soft but now at the M-point.
This can be seen in Fig.~\ref{fig:strain-phonon}(d) for $+16.5$\% strain. This is an
in-plane mode, and thus indicates a structural transition in the 2D monolayer. An optical mode also becomes soft over the entire 2D BZ.
Eigenvectors of these two modes are shown in Fig.\ S6 in SI~\cite{si}. 



  
We calculate the band structures of MCCT by applying biaxial strain in steps of 2~\% on both the
compressive and tensile sides. Fig.~\ref{fig:strain-band} shows the PBE-vdW-DF2 band structure of MCCT under different strain conditions. 
In the unstrained case, MCCT is an indirect band gap semiconductor with a gap of 1.15~eV as mentioned earlier.
Under compressive strain, the indirect nature of the band gap is maintained 
with VBM and CBM still at the $\Gamma$ and $\Delta$ points respectively. However, the band gap drops marginally down to 
1.13~eV at $-4$\% strain. Difference between the direct and indirect gaps is quite small as seen
in panels (a) and (b) in Fig.~\ref{fig:strain-band}.

With tensile strain, the fundamental gap remains indirect between $\Gamma$ and $\Delta$ up to 2\% strain 
(Fig.~\ref{fig:strain-band}-(c)). However, at 4\% strain, while CBM still remains at the $\Delta$-point, the VBM moves to the K-point. 
With further increase in tensile strain from 4\% to 10 \%, the band gap decreases from 1.15~eV to 0.80~eV 
with the VBM still at K-point shown in panels (d)-(g) in Fig.~\ref{fig:strain-band}. At 12 \% strain, the VBM 
shifts to the M-point. The band gap reduces further to 0.69~eV at 12 \%, and to 0.55~eV at 15 \% strain. This can be
seen in panels (h)-(j) in Fig.~\ref{fig:strain-band}.  The energy difference between the valence band states at M and K points 
is always small. Similarly, the energy difference between the conduction band states at these two points is also small. 
As a consequence, the difference between the direct and indirect gaps is small at all strains.
In fact, the valence and conduction bands being rather flat between M and K points, there are lot of states available 
for electron-hole excitations in a narrow energy range. This should make monolayer CCT an attractive material for photovoltaic applications.

\begin{figure}[t]
\includegraphics[scale=0.5]{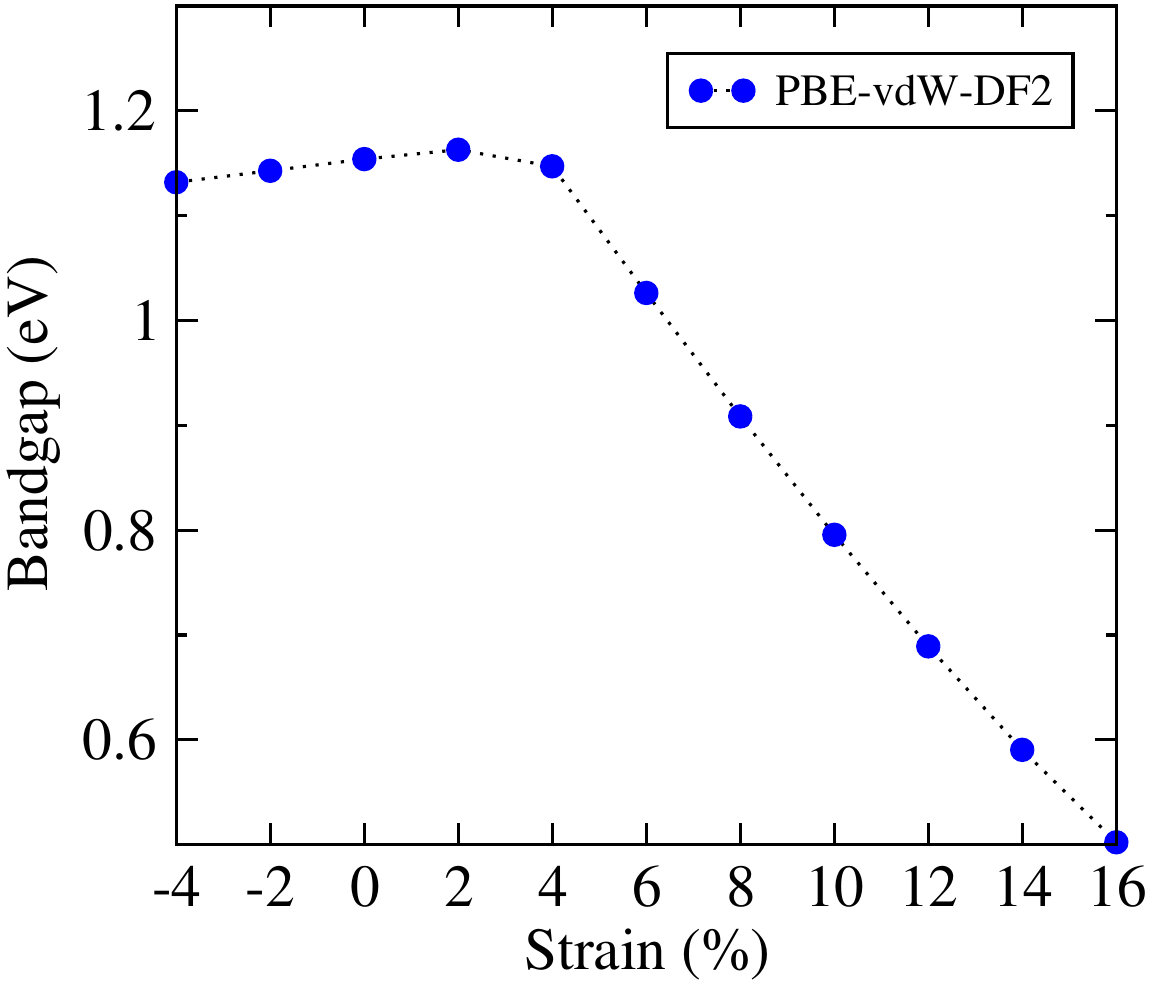} 
 \caption{Variation of band gap of monolayer CCT with biaxial strain.}
 \label{fig:strain-gap}
\end{figure}

 \begin{figure*}[t]
\includegraphics[scale=0.4]{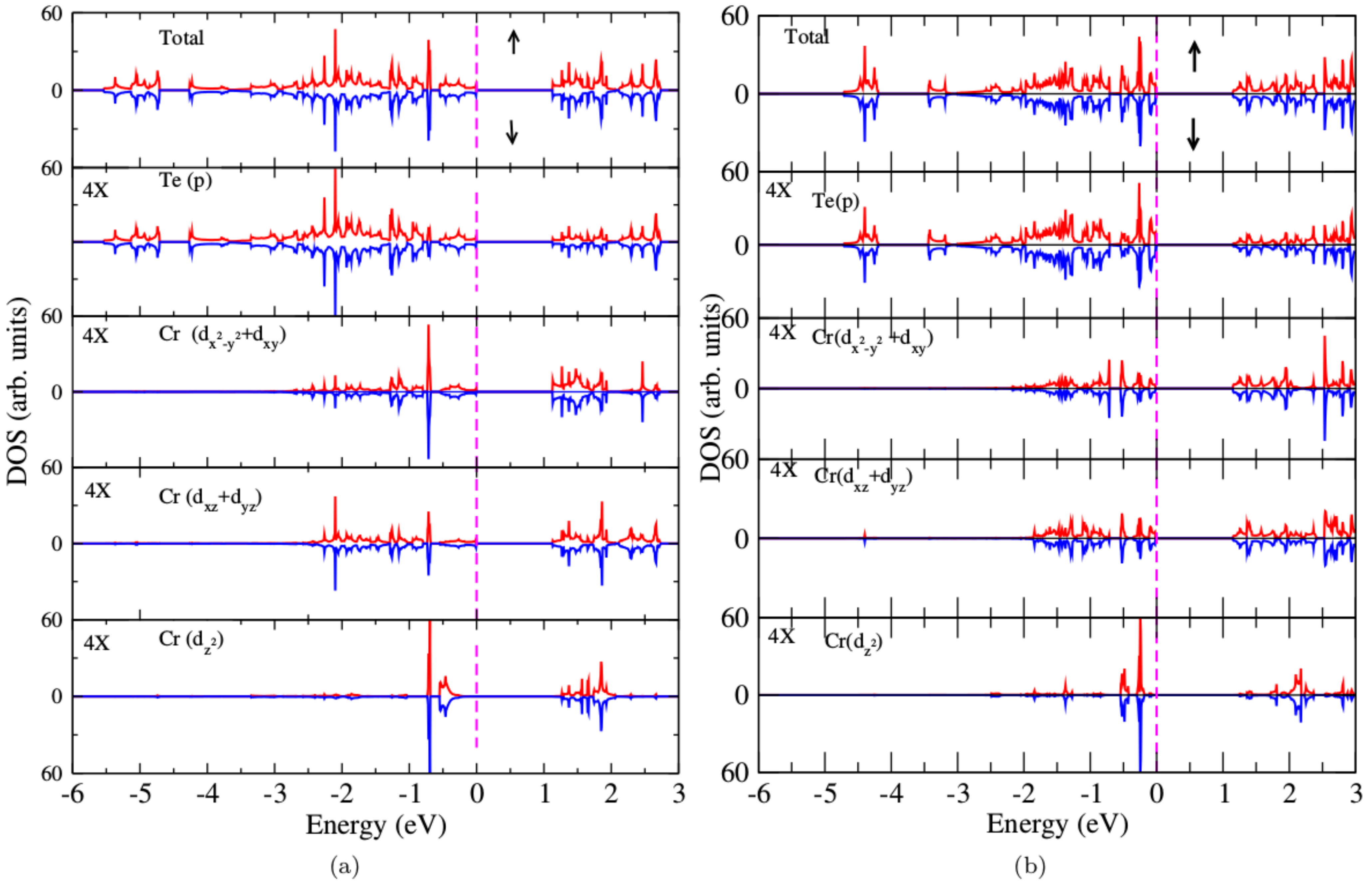}
   \caption{Total, and orbital resolved DOS of monolayer CCT at (a)-4\% strain and (b)+4\% strain.}     
  \label{fig:strain-dos}
 \end{figure*}
 
Variation of the band gap of MCCT with strain is shown in Fig.~\ref{fig:strain-gap}. The band gap changes
very little under compressive strain. It increases marginally between $0-2$\% strain,
but then starts decreasing. Beyond 4\% strain, it decreases sharply all the way up to 16\%. Therefore, 
if our conclusion about structural stability of MCCT is valid in the high strain regime, 
band gap of MCCT can be tuned by as much as 50\% by applying tensile strain. 

In order to get more insights into the electronic structure of MCCT at different strains, we calculated atom and orbital resolved DOS 
at various strains. These quantities at $-4$\%  and $+4$\% strain are shown in Figs.~\ref{fig:strain-dos}(a) \& (b). 
As in the pristine monolayer, the VBM originates primarily from the Te $p$ states, while the CBM has major contribution from the 
Cr $d$ states. A comparison of Figs.~\ref{fig:monolayer-dos} and ~\ref{fig:strain-dos} also show that as the lattice constants of
monolayer CCT increase (tensile strain starting from its $-4\%$ structure),
the contribution of the Cr $d_{x^2-y^2}$, $d_{xy}$, $d_{yz}$ and $d_{zx}$ orbitals to the
states in the range VBM to $-2$~eV decreases. It is the maximum at $-4\%$, and is much lower at $+4\%$. This can
be due to increasing Cr-Te distance with increasing tensile strain. The peak of the Te partial DOS in this energy range
shifts to higher energies. A decrease in the mixing with the Cr $d$ states destabilizes the Te $p$ states somewhat.
At still higher tensile strains (SI)~\cite{si}, new states with contributions from both Cr $d$ and 
Te $p$ orbitals appear $\sim 0.5$~eV above the VBM (Fig.\ S7) forming the new conduction
band edge. These states reduce the band gap as seen Fig.~\ref{fig:strain-gap}.
  


An important question in the context of strain engineering is how robust the AFM order is with respect to applied strain.
Specifically, it is possible that under tensile strain, as the Cr-Cr distance increases, the AFM state may become unstable
towards a FM ground state. In order to check this we calculated energies of both FM and AFM1 ordered states of the Cr
spins under all strain conditions from $-4$\% to $+16$\%. 
It is interesting to note that $\Delta E$
is positive at all strains indicating greater stability of the AFM1 state, and in fact, beyond $\sim 2$\% tensile strain
the AFM1 state gains more stability relative to the FM state.  
In pristine MCCT, the Cr-Te bond length is 2.80 \AA, and Cr-Cr bond length is 3.83 \AA~, with a 
Cr-Te-Cr bond angle of 86.24${\degree}$, as stated earlier. 
With tensile strain, Cr-Te and Cr-Cr bond lengths increase monotonically and reach 3.11 \AA~  and 4.44 \AA~ at +16$\%$ strain. 
The bond angle also increases to $90.89{\degree}$ at $+16$\% strain. With compressive strain, Cr-Te and Cr-Cr bond lengths 
decrease monotonically to reach 2.74 \AA~  and 3.67 \AA~ at $-4$\% strain. The bond angle also reduces to 84.13$\degree$ 
at $-4\%$ strain. Variation of bond lengths and bond angles are given in Table S4 in the SI~\cite{si}.
For comparison, we optimized the structure of monolayer CrSiTe$_3$ with the PBE-vdW-DF2 functional. 
Cr-Cr bond distance in this compound is 4.06 $\AA$  and the Cr-Te-Cr bond angle is $89.97{\degree}$.  
Cr-Te bond length is found to be 2.87 \AA. What is interesting is that the Si compound has a FM ground state with a Cr-Cr
distance of 4.06 \AA, but MCCT remains AFM up to a Cr-Cr distance of 4.44 \AA~ at $+16$\% strain. 

It is puzzling that the AFM1 state becomes more stable with larger tensile strain. Tensile strain monotonically
increases Cr-Cr distance. Therefore, any direct AFM exchange between neighboring Cr atoms must become
weaker. However, as we discussed earlier, the superexchange in the TM tri-chalcogenides are quite
complex. Therefore, one can only speculate that perhaps in CCT it has an AFM character. This issue
requires further careful study.

\section{Conclusion}
\label{conclude}

Using DFT and DFPT calculations we have established that CrCTe$_3$ is a structurally stable compound in the R$\bar{3}$
structure. It turns out to be an AFM semiconductor, a welcome addition to this family. The fundamental band gap is 1.12~eV and
an indirect one  as found in our PBE calculations. The successive layers of the bulk material are rather weakly bound by
van der Waals forces. The cleavage energy is estimated to be 0.24~J/m$^2$, smaller than that of graphene, CrSiTe$_3$ and
CrGeTe$_3$. Monolayers of CCT are also AFM semiconductors with nearly the same indirect gap as the bulk. MCCT
remains structurally stable between 4\% compressive and 16\% tensile biaxial strain. A buckling transition sets in at 4\% 
compressive strain. Phonon softening at $\sim16$\% tensile strain indicates a structural phase transition. Interestingly, the
ground state remains AFM over the entire strain range of $-4$ to $+16$\%. The band gap can be tuned by nearly 50\% by
applying 16\% strain. We hope these results will enthuse experimentalists to synthesize this novel material and to explore
its properties in more detail.

{\bf Acknowledgements} All computations were performed at the HPC cluster facility at HRI
(http://www.hri.res.in/cluster/).


\bibliographystyle{apsrev4-1} 
\bibliography{cct}
\end{document}